\documentclass[twocolumn]{aastex7}

\usepackage{booktabs}
\usepackage{hyperref}
\usepackage{mathtools}
\usepackage{amsmath}
\usepackage{amssymb}
\usepackage{graphicx}
\usepackage{overpic}
\usepackage{natbib}
\usepackage{xcolor}
\usepackage{bm}

\def	\K		{\,{\rm K}}
\def	\mum	{\,{\mu \rm{m}}}
\def \bea {\begin{eqnarray}}
	\def \ena {\end{eqnarray}}
\def	\bB	{\boldsymbol{B}}
\def	\bk	{\boldsymbol{k}}
\def	\cm	{\,{\rm cm}}
\def	\pc	{\,{\rm pc}}
\def	\erg	{\,{\rm erg}}
\def	\km	{\,{\rm km}}

\def	\yr	{\,{\rm yr}}

\providecommand{\um}{\,\mu\mathrm{m}}
\providecommand{\Lsun}{L_{\odot}}

\providecommand{\NHsn}{N_{\mathrm{\rm H}}}
\providecommand{\Td}{T_{\mathrm{d}}}
\providecommand{\Tgas}{T_{\mathrm{gas}}}

\providecommand{\nH}{n_{\mathrm{H}}}
\providecommand{\Smax}{S_{\mathrm{max}}}

\providecommand{\adisr}{a_{\mathrm{disr}}}
\providecommand{\afast}{a_{\mathrm{fast,min}}}
\providecommand{\fHiISRF}{f_{\mathrm{hiJ}}^{\mathrm{ISRF}}}
\providecommand{\fhif}{f_{\mathrm{hiJ}}^{\mathrm{fast}}}
\providecommand{\Rhigh}{R_{\mathrm{high\text{-}J}}}
\providecommand{\Rlow}{R_{\mathrm{low\text{-}J}}}
\providecommand{\pem}{p_{\mathrm{em}}}
\providecommand{\pext}{p_{\mathrm{ext}}}
\providecommand{\Ncl}{N_{\mathrm{cl}}}
\providecommand{\RVsn}{R_{V}}

\newcommand{\TransRAT}{\textsc{TransRAT}}
\newcommand{\BRAT}{B\text{-}\mathrm{RAT}}
\newcommand{\kRAT}{k\text{-}\mathrm{RAT}}
\newcommand{\Bv}{\boldsymbol{B}}
\newcommand{\kv}{\boldsymbol{k}}
\newcommand{\paperii}{Paper~II}

\graphicspath{
  {./}
  {./figures/}% <-- copy the MC/amax_1um/figures_SNeII_... directories here for submission
  {/Users/thiemhoang/Dropbox/Desktop/4.Projects/DUST_RAT_Transients/Transient_MC/output/MC/amax_0p5um/}
  {/Users/thiemhoang/Dropbox/Desktop/4.Projects/DUST_RAT_Transients/Papers/working/SNe-MCs/}
}

\begin{document}

\title{Time-Domain Dust Astrophysics. I. Polarization Flares, Polarization-Angle Reverberation, and Fossil Imprints in Supernova-Illuminated Clouds}

\author{Thiem Hoang}
\affiliation{Korea Astronomy and Space Science Institute, Daejeon 34055, Republic of Korea}
\affiliation{Department of Astronomy and Space Science, University of Science and Technology, 217 Gajeong-ro, Yuseong-gu, Daejeon, 34113, Republic of Korea}

\email{thiemhoang@kasi.re.kr}

%\author{Co-author}
%\affiliation{Co-author Affiliation}
%\email{coauthor@email.com}

\begin{abstract}
Cosmic transients can enhance the local radiation field on timescales of days to months, rendering the steady-state radiative torque (RAT) paradigm of grain alignment inapplicable. Using the time-domain \TransRAT\ framework, which self-consistently follows grain heating, alignment, rotational disruption, and switching of the alignment axis between the magnetic field ($\Bv$) and the radiation direction ($\kv$),
we predict the time-dependent dust polarization of a dense cloud, represented by a homogeneous one-zone model, illuminated by a Type~IIP supernova over a range of source--cloud zone distances. We identify four key signatures. First, for $D\lesssim1\pc$, a \emph{polarization flare} develops within days to weeks, marked by sharp increases in both thermal dust polarization and extinction-polarization efficiency; this is followed by a \emph{polarization dip} as radiative torque disruption (RAT-D) destroys the large aligned grains. Second, the peak wavelength of extinction polarization, $\lambda_{\max}$, shifts blueward as the minimum aligned-grain size decreases, providing a diagnostic largely independent of magnetic-field geometry. Third, the transition from $\BRAT$ to $\kRAT$ produces an abrupt \emph{polarization-angle rotation} of up to $45^{\circ}$ in our fiducial geometry. Fourth, as the transient fades, the return to $\BRAT$ generates a \emph{polarization-angle reverberation} governed by Larmor precession. This reverberation is the most sensitive probe of grain magnetism, with superparamagnetic grains recovering more rapidly than paramagnetic grains. At $D>1\pc$, SN-induced polarization properties can persist long after the radiation has faded. These {\it fossil imprints} offer the most practical near-term observational test: clouds near supernova remnants younger than the relaxation timescale of $\sim10\,t_{\rm gas}$ with $t_{\rm gas}$ gas damping time, should exhibit elevated polarization and blueshifted $\lambda_{\max}$ today. Time-dependent dust polarization can therefore probe dust physics, dust properties, and pristine pre-shock magnetic fields in real time, while fossil imprints preserve a record of past explosions—establishing dust polarization as a new messenger of time-domain astrophysics.
\end{abstract}


\section{Introduction}
\label{sec:intro}

Polarization of starlight and thermal dust emission produced by aligned grains provides a powerful probe of magnetic fields in environments ranging from the diffuse interstellar medium (ISM) and molecular clouds (MCs) to protostellar cores and disks \citep{Andersson.2015,PattleFissel.2019}. The leading theory of interstellar grain alignment is based on radiative torques (RATs) \citep{Dolginov.1976,DraineWein.1997,LazHoang.2007,HoangLaz.2008}. Under the interstellar radiation field (ISRF) or protostellar radiation, Larmor precession generally causes grains to align with the magnetic field, a regime known as $\BRAT$ alignment \citep{Hoangetal.2022}. In sufficiently strong and anisotropic radiation fields, radiative precession can instead dominate, causing grains to align with the radiation direction $\kv$ through $\kRAT$ alignment \citep{LazHoang.2007,Hoangetal.2022}. The competition between these two alignment axes depends primarily on the radiation field and the magnetic susceptibility of the grains \citep{Hoangetal.2022,Hoang.2025}. Together with magnetically enhanced RAT alignment \citep{HoangLaz.2016} and radiative torque disruption \citep[RAT-D;][]{Hoangetal.2019}, the RAT paradigm has been tested extensively against starlight and far-IR/submillimeter polarization observations (see reviews by \citealt{Andersson.2015,TramHoang.2022}). These tests, however, have largely probed steady environments in which grain alignment has already approached equilibrium \citep{HoangLaz.2008,HoangLaz.2016}. They do not test a central dynamical prediction of RAT theory: under intense radiation, grains can undergo \emph{fast alignment} on timescales shorter than the gas-damping time \citep{LazHoang.2007,LazHoang.2019,Hoang.2025}. Until the development of the framework used here, this process had not been incorporated into a fully time-dependent polarization model and remained observationally untested.

%Steady-state RAT polarization has been modeled in a variety of environments using DustPOL-py \citep{Lee.2019,TramHoang.2022} and POLARIS \citep{Reissl.2016}.

It is noted that the basic theory of RAT alignment \citep{LazHoang.2007,Hoang.2025} and RAT-based polarization modeling \citep{Hoangetal.2014,LeeHoang.2020} assume that the radiation field is constant or varies slowly compared with the characteristic alignment timescales . This assumption breaks down around cosmic transients---including supernovae (SNe), gamma-ray bursts (GRBs), novae, and tidal disruption events (TDEs)---whose radiation energy density can increase by many orders of magnitude over days to months (e.g., \citealt{Hinkle.2025}). Dusty transient environments are common. Core-collapse SNe (CCSNe), long GRBs, and magnetars originate from massive stars born in giant molecular clouds, and many supernova remnants (SNRs) are observed near or interacting with molecular material \citep{Zhou.2023,Reach.2024}. Runaway massive stars can explode within $1\pc$ of unrelated clouds \citep{Dincel.2026}, while dense clumps produced by massive-star winds can survive within $\lesssim10\pc$ of the explosion site \citep{slane.2015supernova-c88}. In each of these environments, dust initially aligned by the ambient ISRF can be abruptly exposed to an intense and rapidly evolving radiation field.

Such time-dependent illumination introduces several coupled effects that are separately treated in steady-state models: transient grain heating that alters the magnetic susceptibility and Larmor precession \citep{Hoang.2026}; fast RAT alignment on timescales comparable to the radiative-precession time $\tau_k$ \citep{LazHoang.2007,LazHoang.2019,Hoang.2025}; irreversible rotational disruption of large grains by RAT-D \citep{Hoangetal.2019}; and switching of the alignment axis between $\Bv$ and $\kv$ (Figure~\ref{fig:BRAT_kRAT}). Previous studies modeled grain alignment and disruption under SN and GRB illumination while assuming that grains retained their pre-existing alignment axis \citep{Gianghoang.2020,Hoanggiang.2020}. A unified framework that evolves heating, alignment, disruption, and alignment-axis switching self-consistently and maps their coupled evolution into time-dependent observables has therefore been lacking.

In a companion paper (\citealt{Hoang.2026klf}, hereafter \paperii), we develop \TransRAT\ (Transient Radiative Torque), a general time-domain framework for modeling grain heating, grain alignment, disruption, and their observational signatures under an arbitrary bolometric light curve. In this paper, we present its central polarimetric predictions for a dense cloud represented by a homogeneous one-zone model illuminated by a Type~IIP SN and key polarization signatures. The rest of the paper is organized as follows. Section~\ref{sec:model} summarizes the physical model and model setup, and Section~\ref{sec:results} presents the time-dependent polarization signatures. Section~\ref{sec:discussion} discusses in detail the main results, their physical implications, and prospects for testing the predicted signatures with real-time and archaeological observations.

\section{Model setup and summary}
\label{sec:model}
We consider a dense cloud illuminated by a Type~IIP SN with its time-dependent bolometric luminosity, $L(t)$. To isolate the time dependent dust physics, we approximate the illuminated cloud by a homogeneous, single-phase, one-zone model instead of treating three-dimensional cloud structure. Let $D$ be the distance between the cloud zone and the transient source. The arrival time of the SN radiation at the cloud zone
\begin{equation}
	t_r=t-\frac{D}{c}
\end{equation}
is the retarded time, with $t=0$ denoting the explosion and $t_r=0$ the arrival of the first SN light at the cloud.

Given $L(t_r)$, the \TransRAT\ framework self-consistently evolves the coupled thermal, rotational, and alignment states of the dust. It first calculates grain heating and sublimation, including the time-dependent grain temperature $\Td(t_r)$. It then solves the RAT spin-up equation of motion for the grain angular velocity and identifies the irreversible RAT-D disruption interval, $[\adisr^{\min},\adisr^{\max}]$, within which centrifugal stress exceeds the grain tensile strength. %The disrupted grains are removed from the evolving grain-size distribution.

\begin{figure}
	\includegraphics*[width=0.5\textwidth]{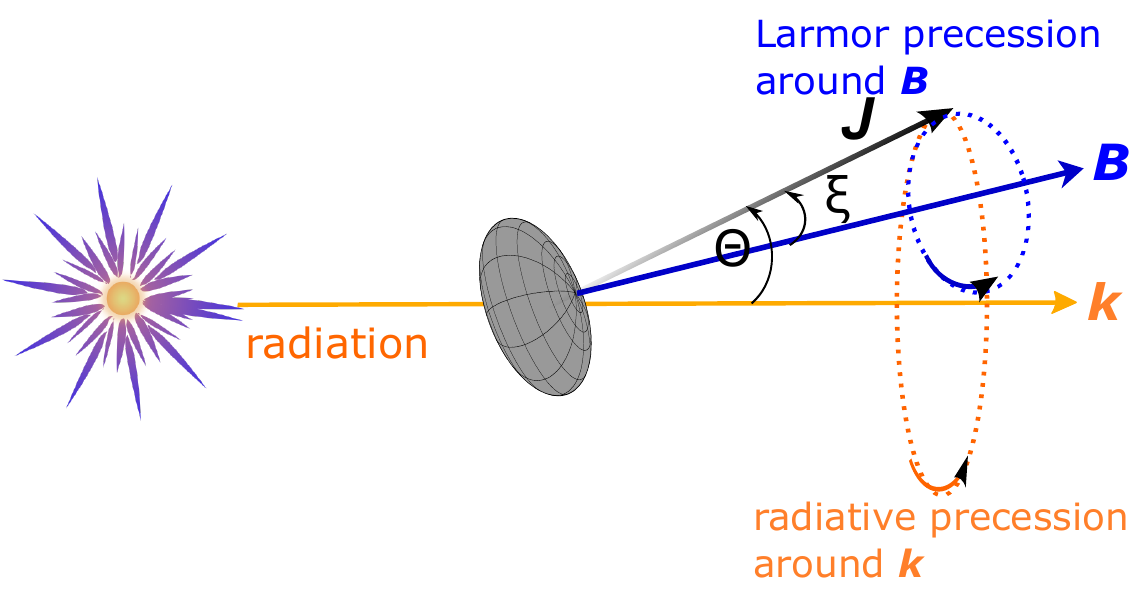}
	\caption{Schematic illustration of the two possible alignment axes of grains exposed to a transient, set by the competition between Larmor precession around the magnetic field $\Bv$ and radiative precession around the radiation direction $\kv$. Before the transient, Larmor precession dominates and grains align with $\Bv$ ($\BRAT$ alignment). The intense transient radiation accelerates radiative precession, and once it outpaces Larmor precession, the alignment axis switches from $\Bv$ to $\kv$ ($\kRAT$ alignment).}
	\label{fig:BRAT_kRAT}
\end{figure}

The alignment axis is determined by the competition between Larmor and radiative precession, which selects either $\BRAT$ or $\kRAT$ alignment (Figure~\ref{fig:BRAT_kRAT}). This competition is evaluated using the temperature-dependent magnetic susceptibility $\chi(\Td)$ for five magnetic models: paramagnetic (PM) grains; superparamagnetic (SPM) grains containing clusters of $\Ncl=10^{2}$, $10^{3}$, or $10^{4}$ iron atoms; and SPM grains with a power-law distribution of cluster sizes. \TransRAT\ also evolves the grain populations at low-$J$ and high-$J$ rotational attractors separately, where $J$ denotes the grain angular momentum. This distinction is essential because the two populations can have different alignment axes, and only grains at high-$J$ attractors undergo rotational disruption.

The resulting time-dependent dust state is passed to the \texttt{DustPOL\_py} polarized radiative-transfer engine \citep{Tram.2023} to predict extinction curves, dichroic extinction polarization, and polarized thermal emission. The complete formulation, validation against the analytically tractable $\bk\parallel\bB$ configuration, and comprehensive predictions for extinction, thermal-emission spectral energy distributions (SEDs), and spinning-dust emission are presented in \paperii.

We apply the \TransRAT\ framework to a dense cloud illuminated by a Type~IIP SN. The adopted light curve has a luminosity normalization $L_{\max}=5.8\times10^{9}\Lsun$ and reaches a peak luminosity of $\simeq0.3\,L_{\max}\approx1.7\times10^{9}\Lsun$. It consists of a 10-day rise, a photospheric plateau, and a $^{56}$Co radioactive-decay tail (see \citealt{Dastidar.2018,Hoangetal.2019}). The fiducial cloud has a gas density $\nH=10^{4}\cm^{-3}$, an initial gas temperature $\Tgas=20\K$, a magnetic field strength $B=100\,\mu{\rm G}$. The grain tensile strength is taken to be $\Smax=10^{9}\erg\cm^{-3}$ for small {\it compact} grains of $a<0.1\mu$m and $\Smax=10^{7}\erg\cm^{-3}$ for larger grains. We adopt the \texttt{astrodust+PAH} grain model \citep{DraineHensley.2023}, with grain sizes spanning $a_{\min}=3.5\,\AA$ to $a_{\max}=0.5\mum$. The oblate spheroidal shape with axial ratio $s=1.4$ is assumed. We adopt a line-of-sight depth of $0.01\pc$, equal to the width of the one-zone region, corresponding to a hydrogen column density $\NHsn\simeq3\times10^{20}\cm^{-2}$. We assume thermal coupling between the gas and dust after illumination, such that $\Tgas=\Td(t_r)$ for $t_r>0$. The angle between the radiation direction and the magnetic field is set to $\psi=45^{\circ}$. We consider the one-zone cloud model located at various distances from the transient source
\begin{equation}
	D=\{0.01,\,0.1,\,1,\,5,\,10,\,20\}\pc
\end{equation}
from the SN. 

For the extinction-polarization calculation, we consider background starlight transmitted through the cloud and place both $\Bv$ and $\kv$ in the plane of the sky. In this geometry, $\psi$ is also the difference between their projected position angles. We separately consider the occulting geometry in which the cloud lies along the sightline to the SN. In that case, $\kv$ points along the line of sight, and $\kRAT$-aligned grains produce no dichroic nor emission polarization. Wherever grains retain $\BRAT$ alignment throughout the transient---as occurs for the distant clouds at $D\geq5\pc$---the same temporal behavior also applies to the polarization of the SN light in the occulting geometry, but with its amplitude reduced by the projection factor $\cos^{2}\gamma$ with $\gamma$ the angle between $\kv$ and the plane of the sky.

Before the explosion, grains are aligned by the ambient ISRF with the dimensionless radiation strength, defined as the local radiation energy density normalized to the ISRF value, $U=0.1$ typical for dense clouds. We adopt the typical values for the high-$J$ fraction $\fHiISRF=0.5$ and Rayleigh reduction factors $(\Rlow,\Rhigh)=(0.1,1.0)$ (see e.g. \citealt{Hoang.2025}).

The full evolution of the quantities controlling the polarization response---including the radiation strength, dust temperature, alignment and disruption sizes, and effective alignment efficiency---is presented in \paperii. As the SN radiation reaches the cloud, the dimensionless radiation strength $U$ increases by approximately 4--10 orders of magnitude depending on $D$. The enhanced radiation heats the grains and can sublimate dust closest to the SN. Simultaneously, stronger RATs reduce the fast-alignment threshold $\afast$, allowing grains as small as $\sim0.01\mum$ to align within days. The disruption size range $[\adisr^{\min},\adisr^{\max}]$ broadens with time as large grains at high-$J$ attractors are irreversibly fragmented into smaller grains. Finally, radiative precession can overtake Larmor precession for grains with insufficient magnetic susceptibility to maintain $\BRAT$ alignment, driving the transition from $\Bv$ to $\kv$ \citep{Hoangetal.2022,Hoang.2026}. Together, these coupled processes produce the time-dependent polarization signatures presented in Section~\ref{sec:results}.

\section{Results: time-dependent polarization signatures}
\label{sec:results}

\subsection{The polarization flare and polarization dip}
\label{sec:flare}

\begin{figure*}
	\centering
	\begin{overpic}[width=0.99\textwidth,percent]{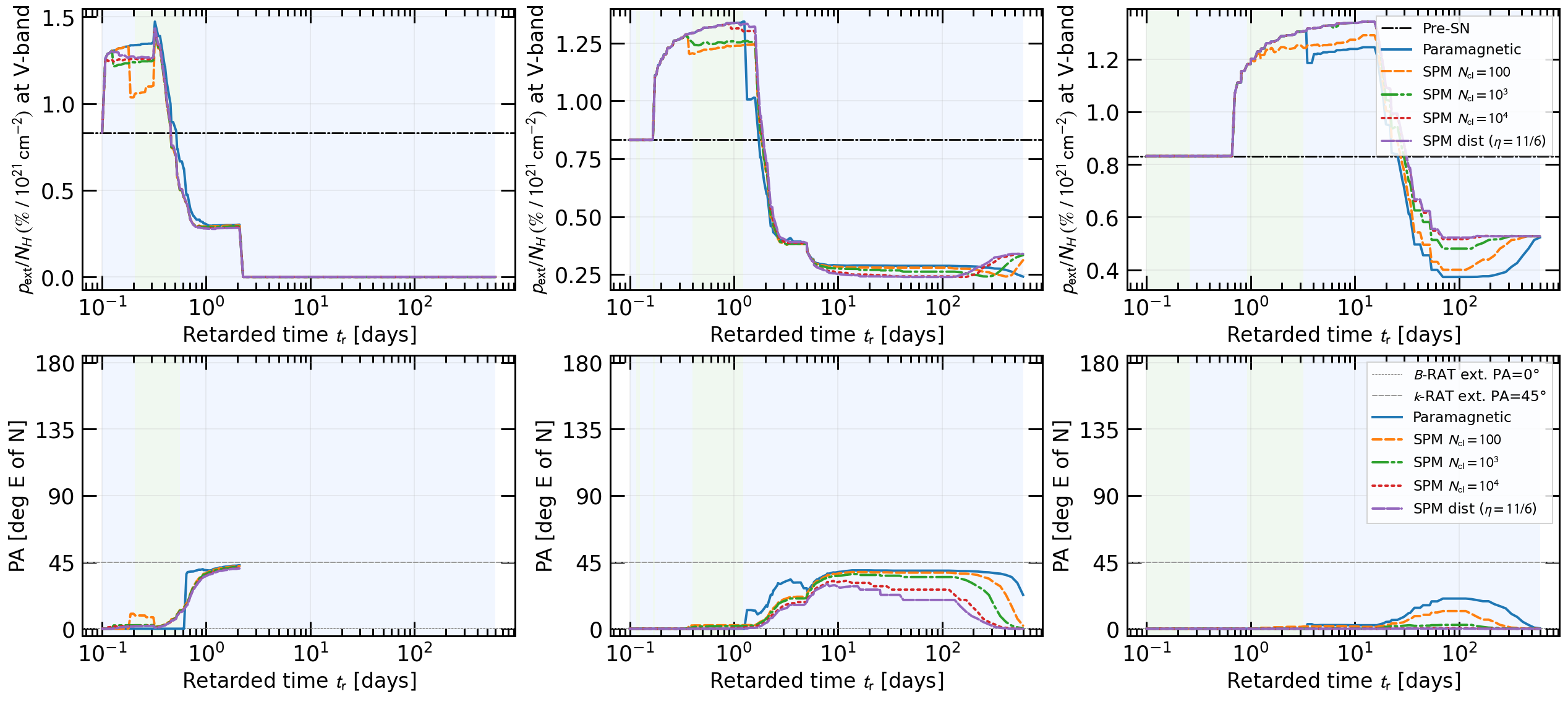}
		% labels placed inside the top-left corner of each panel's plot box
		\put(8.5,28){\textbf{(a)}}\put(41.5,28){\textbf{(b)}}\put(74.5,28){\textbf{(c)}}
		\put(8.5,7){\textbf{(d)}}\put(41.5,7){\textbf{(e)}}\put(74.5,7){\textbf{(f)}}
	\end{overpic}
	\caption{Polarization light curves of background starlight extincted by nearby SN-illuminated clouds at $D=0.01$, $0.1$, and $1\pc$ (left to right columns), for the fiducial geometry $\psi=45^{\circ}$ (the angle between $\Bv$ and $\kv$ in the plane of the sky): dichroic polarization efficiency $\pext(V)/N_{\rm H}$ (upper panels, a--c) and polarization angle (lower panels, d--f). Lines show the five magnetic-susceptibility cases, from PM to SPM with increasing iron-cluster size $\Ncl$. The polarization flare--dip cycle and the $45^{\circ}$ angle rotation occur earlier and more strongly for closer clouds and for lower magnetic susceptibility.}
	\label{fig:pext_mag_Bk45}
\end{figure*}

\begin{figure*}
	\centering
	\begin{overpic}[width=0.99\textwidth,percent]{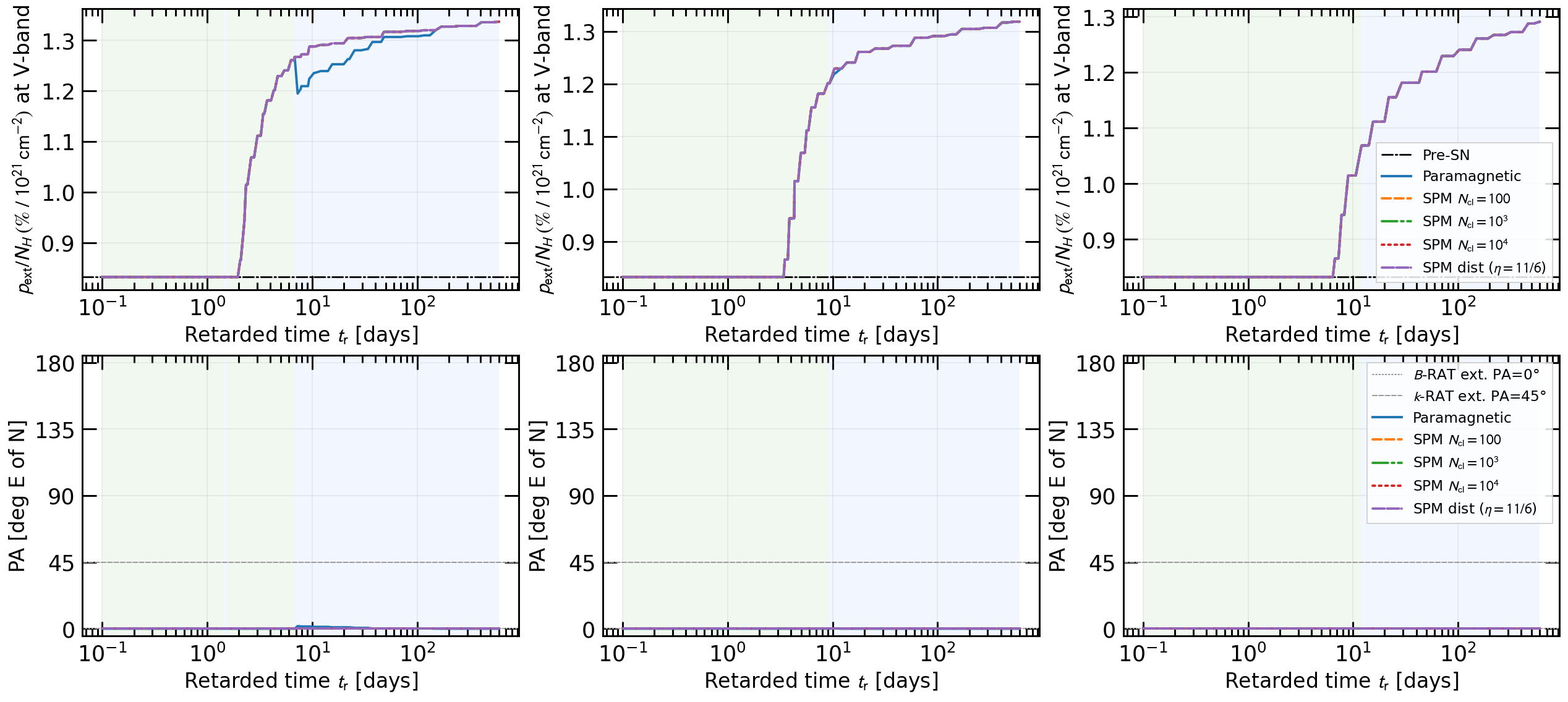}
		% labels placed inside the top-left corner of each panel's plot box
		\put(8.5,28){\textbf{(a)}}\put(41.5,28){\textbf{(b)}}\put(74.5,28){\textbf{(c)}}
		\put(8.5,7){\textbf{(d)}}\put(41.5,7){\textbf{(e)}}\put(74.5,7){\textbf{(f)}}
	\end{overpic}
	\caption{Same as Figure~\ref{fig:pext_mag_Bk45}, but for distant cloud zones at $D=5$, $10$, and $20\pc$ (left to right columns). The flare is slower, no dip occurs because RAT-D is inefficient, and the polarization angle remains essentially constant.}
	\label{fig:pext_mag_Bk45_R1-10}
\end{figure*}

\begin{figure*}
	\centering
	\begin{overpic}[width=0.99\textwidth,percent]{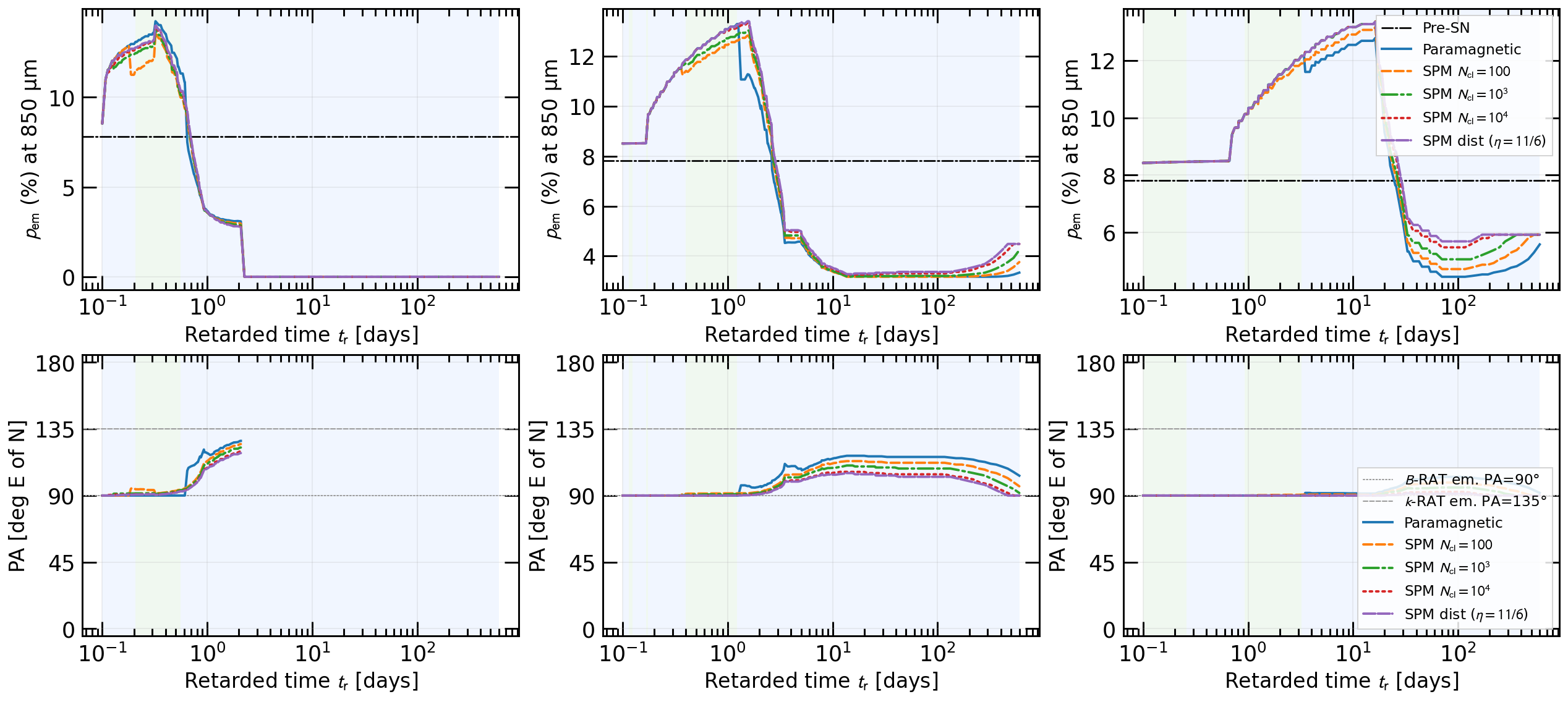}
		% labels placed inside the top-left corner of each panel's plot box
		\put(8.5,29){\textbf{(a)}}\put(41.5,29){\textbf{(b)}}\put(74.5,29){\textbf{(c)}}
		\put(8.5,7){\textbf{(d)}}\put(41.5,7){\textbf{(e)}}\put(74.5,7){\textbf{(f)}}
	\end{overpic}
	\caption{Same as Figure~\ref{fig:pext_mag_Bk45}, but for the thermal dust emission: polarization fraction $\pem(850\mum)$ (upper panels, a--c) and polarization angle (lower panels, d--f) for nearby cloud zones at $D=0.01$, $0.1$, and $1\pc$. The abrupt change of the polarization angle marks the $\BRAT\to\kRAT$ alignment-axis switch, which occurs for PM and low-$\Ncl$ SPM grains but is suppressed at high magnetic susceptibility.}
	\label{fig:pem_mag_Bk45}
\end{figure*}

\begin{figure*}
	\centering
	\begin{overpic}[width=0.99\textwidth,percent]{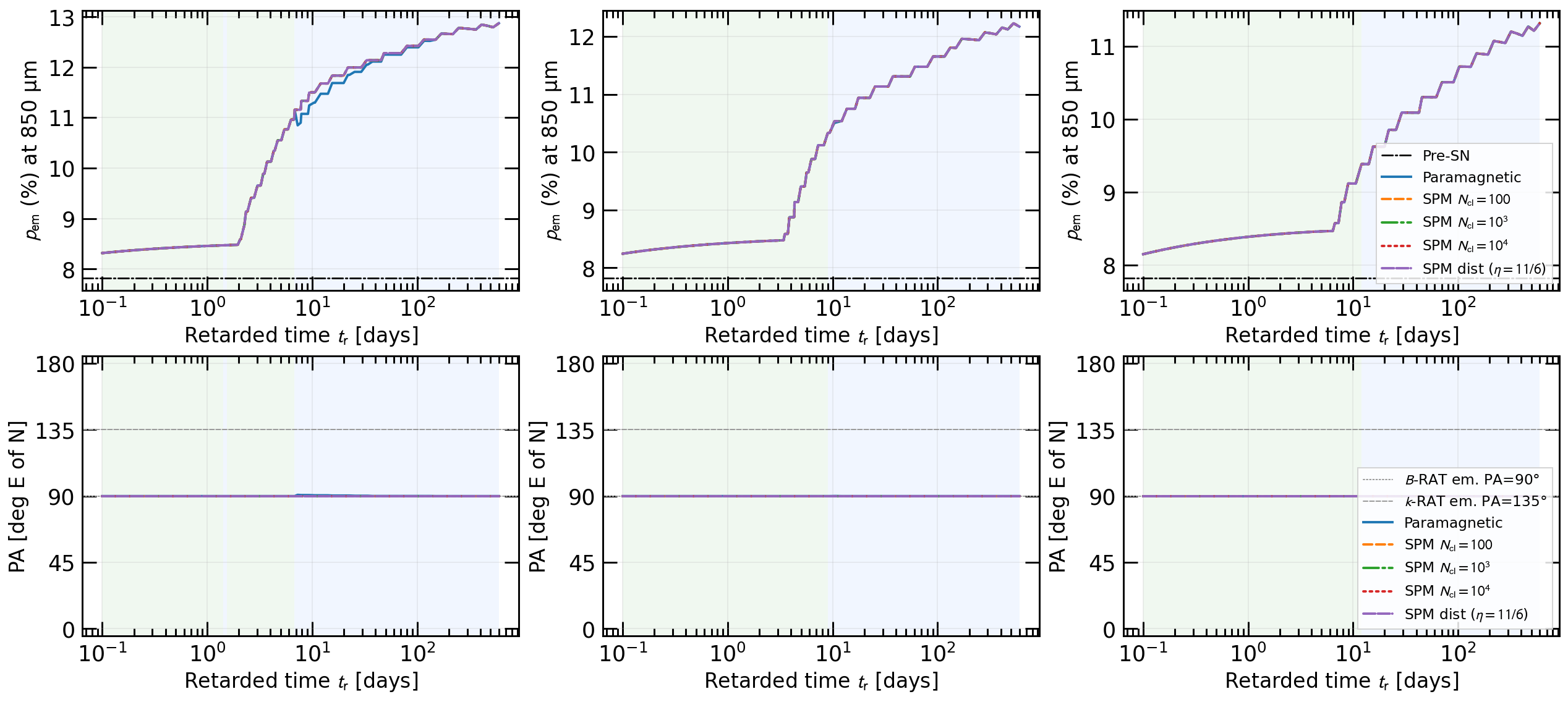}
		% labels placed inside the top-left corner of each panel's plot box
		\put(8.5,41){\textbf{(a)}}\put(41.5,41){\textbf{(b)}}\put(74.5,41){\textbf{(c)}}
		\put(8.5,19){\textbf{(d)}}\put(41.5,19){\textbf{(e)}}\put(74.5,19){\textbf{(f)}}
	\end{overpic}
	\caption{Same as Figure~\ref{fig:pem_mag_Bk45}, but for distant cloud zones at $D=5$, $10$, and $20\pc$ (left to right columns). The emission polarization rises gradually and saturates without a dip, and no $\BRAT\to\kRAT$ switch occurs.}
	\label{fig:pem_mag_Bk45_R1-10}
\end{figure*}

Figures~\ref{fig:pext_mag_Bk45} and \ref{fig:pext_mag_Bk45_R1-10}, panels (a)--(c), show the time-varying extinction-polarization efficiency at $V$ band, $\pext(V,t_r)/N_{\rm H}$, for background starlight transmitted through nearby and distant cloud zones, respectively. Figures~\ref{fig:pem_mag_Bk45} and \ref{fig:pem_mag_Bk45_R1-10} show the corresponding thermal-emission polarization at $850\mum$, $\pem(t_r)$, over similar ranges of SN-cloud distance.

For clouds at $D\lesssim1\pc$, both observables exhibit a characteristic rise--peak--decline evolution relative to their quiescent baselines: a \emph{polarization flare} followed by a \emph{polarization dip}. Within days to weeks of the explosion, RATs rapidly align an increasing fraction of the grain population as the fast-alignment threshold $\afast$ shifts to smaller sizes. Consequently, $\pem$ rises steeply from its pre-SN value of $\simeq 7.8\%$ (already $\simeq8.5\%$ at the first plotted epoch $t_{r}=0.1$\,d, when the grains are mildly heated) and can reach $\simeq13$--$14\%$ at $D\lesssim1\pc$. The polarization reaches its maximum when the fast-aligned fraction saturates, producing the flare. Its amplitude is controlled primarily by the fraction of fast-aligned grains that reach high-$J$ attractors, $\fhif$. Subsequently, RAT-D destroys the large grains that dominate the polarized signal, causing the polarization to decline over tens to hundreds of days. This decline defines the polarization dip, whose onset occurs earlier and more abruptly at smaller $D$ and whose depth reflects the disruption efficiency. Because RAT-D irreversibly modifies the grain-size distribution, the polarization deficit persists after the transient radiation fades, unlike the short-lived flare. The complete flare--dip cycle lasts $\sim0.2$--$15$ days at $D=0.1\pc$ and $\sim0.7$--$100$ days at $D=1\pc$. In the extreme radiation field at $D=0.01\pc$, thermal sublimation instead destroys the grains and drives the polarization to zero within $\sim2$ days. 

The extinction polarization $\pext(V)$ exhibits a similar flare--dip evolution with comparable amplitudes for near clouds (Figure \ref{fig:pext_mag_Bk45}). At $D\geq5\pc$, the radiation field is too weak to produce significant RAT-D. The polarization therefore rises as fast alignment proceeds and then approaches saturation, without developing a pronounced disruption-driven dip (Figures~\ref{fig:pext_mag_Bk45_R1-10} and \ref{fig:pem_mag_Bk45_R1-10}). The timing and shape of the flare can thus constrain the SN-cloud distance, while the presence or absence of a subsequent dip diagnoses whether the radiation field is sufficiently strong to trigger rotational disruption.

The two branches of the light curve probe complementary physics: the rising branch traces fast alignment, whereas the declining branch traces rotational disruption. Their combination therefore tests both dynamical predictions within a single event. A control calculation in which most grains initially occupy low-$J$ attractors ($\fHiISRF=0.01$) changes the quiescent polarization baseline but leaves the transient rise and decline nearly unchanged (\paperii). The time-variable polarization component is therefore a robust prediction that is relatively insensitive to the uncertain pre-SN alignment state and can help distinguish RAT-driven variability from scattered-light echoes.

\subsection{Peak wavelength of extinction polarization}
\label{sec:lambdamax}
\begin{figure*}
	\centering
	\begin{overpic}[width=0.99\textwidth,percent]{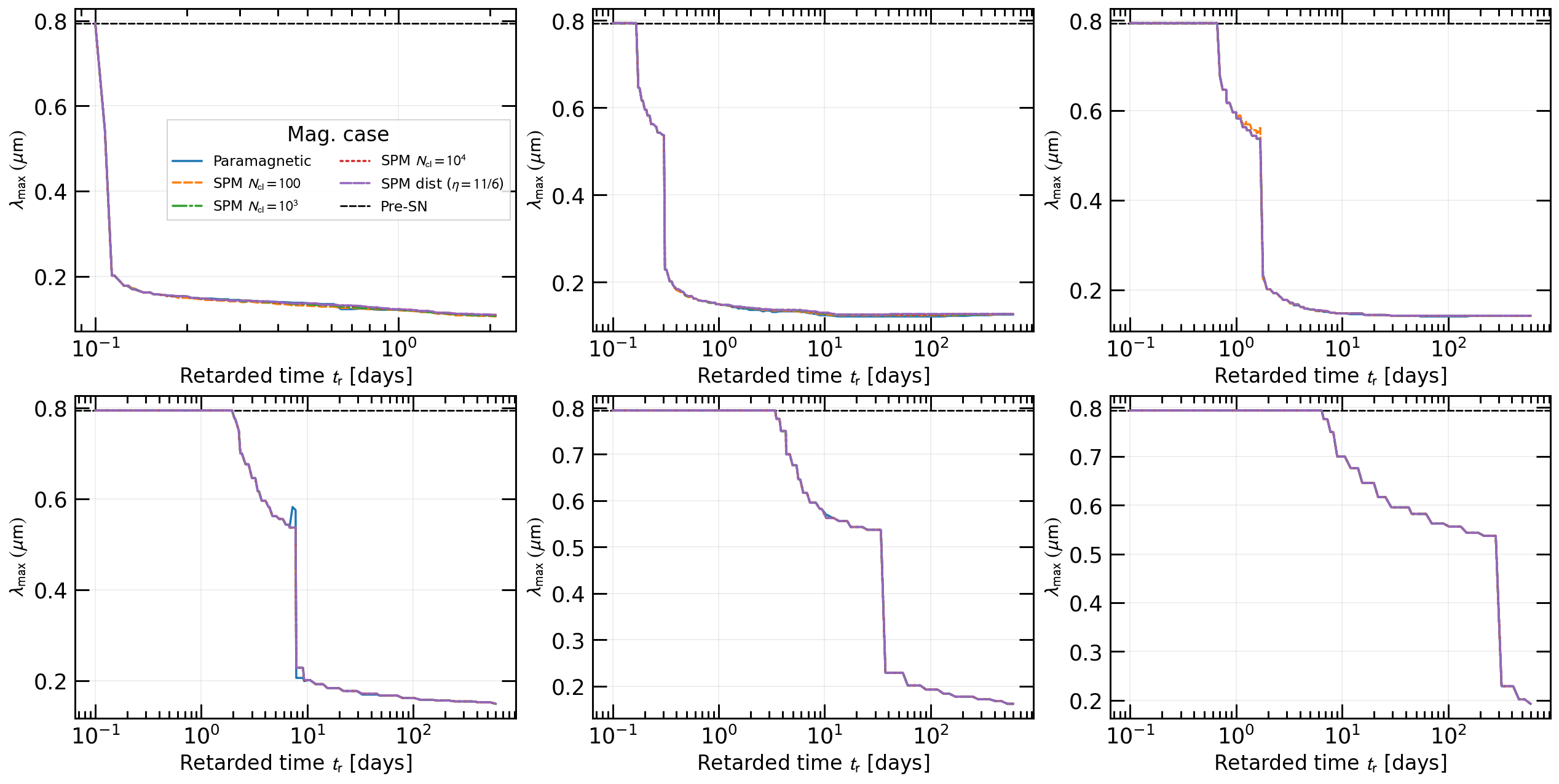}
		% labels placed inside the top-left corner of each panel's plot box
		\put(8.5,46){\textbf{(a)}}\put(41.5,46){\textbf{(b)}}\put(74.5,46){\textbf{(c)}}
		\put(8.5,22){\textbf{(d)}}\put(41.5,22){\textbf{(e)}}\put(74.5,22){\textbf{(f)}}
	\end{overpic}
	\caption{Time evolution of the peak wavelength $\lambda_{\max}$ of the extinction polarization spectrum for the dust cloud at different distances from the SN (panels a--f: $D=0.01$--$20\pc$) and the five magnetic-susceptibility cases ($\psi=45^{\circ}$). The rapid blueward drift of $\lambda_{\max}$ from its quiescent value $\approx 0.8\mum$ traces the decreasing minimum size of aligned grains driven by fast alignment.}
	\label{fig:lambda_peak_mag_Bk45}
\end{figure*}

The wavelength dependence of extinction polarization provides information complementary to the polarization amplitude. Figure~\ref{fig:lambda_peak_mag_Bk45}, panels (a)--(f), shows the evolution of the peak wavelength $\lambda_{\max}$ for each distance and magnetic configuration. At $t_r=0$, before the transient illumination, the polarization spectrum peaks at $\lambda_{\max}\simeq0.8\um$. This value is larger than the diffuse-ISM value of $\simeq0.55\um$ because reduced interstellar radiation and increased gas collisions in the dense cloud allows only the larger grains to remain aligned \citep{Hoang.2021}.

As the SN radiation intensifies, fast alignment extends progressively toward smaller grains, with $\afast$ decreasing to $\sim0.01\um$ (\paperii). The resulting increase in the aligned small-grain population drives a rapid blueward shift of $\lambda_{\max}$ over days to weeks. For $D=0.01\pc$, the curve is shown only up to $\sim2$ days, after which thermal sublimation destroys the dust. The temporal evolution of $\lambda_{\max}$ therefore directly traces the changing fast-alignment threshold size.

Unlike the polarization amplitude, $\lambda_{\max}$ is largely insensitive to the magnetic-field inclination and dust column. A time-variable $\lambda_{\max}$ correlated with the SN illumination history thus provides a geometry-independent diagnostic of time-dependent grain alignment. It can be measured through multiband imaging polarimetry or spectropolarimetry of background stars seen through the illuminated cloud, or directly from the SN when the cloud lies along the line of sight.

% Because the enhanced alignment relaxes only over $\tau_{\rm relax}\sim10$--$100\,t_{\rm gas}$, $\lambda_{\max}$ may remain blueshifted long after the transient fades, leaving a spectral fossil imprint complementary to the change in $\RVsn$ (\paperii).

\subsection{Polarization-angle rotation driven by the $\BRAT\to\kRAT$ transition}
\label{sec:angle}

The polarization angle provides a direct diagnostic of the grain-alignment axis, distinguishing alignment with the magnetic field ($\Bv$) from alignment with the radiation direction ($\kv$). As the radiation field intensifies, radiative precession overtakes Larmor precession for grains with low magnetic susceptibility, allowing their alignment axis to switch from $\Bv$ to $\kv$. The observable angle rotation, however, emerges only after RAT-D removes the large high-$J$ grains whose $\BRAT$ alignment initially dominates the polarized signal. The remaining $\kRAT$-aligned population then becomes dominant, rotating the net polarization angle.

For our fiducial geometry with $\psi=45^{\circ}$, panels (d)--(f) of Figures~\ref{fig:pext_mag_Bk45} and \ref{fig:pem_mag_Bk45} show a polarization-angle rotation of up to $45^{\circ}$ in both extinction and thermal-emission polarization on timescales of days to weeks for clouds at $D\lesssim1\pc$. The amplitude of the rotation decreases with increasing $D$: in extinction polarization the angle reaches the full $\kRAT$ value of $\simeq45^{\circ}$ at $D=0.01\pc$ and $\simeq43^{\circ}$ at $D=0.1\pc$, but only $\simeq25^{\circ}$ at $D=1\pc$, where a smaller fraction of the polarized signal is carried by $\kRAT$-aligned grains. The emission polarization angle behaves similarly, rotating from its $\BRAT$ value of $90^{\circ}$ by $\simeq35^{\circ}$ at $D=0.01\pc$ and $\simeq25^{\circ}$ at $D=0.1\pc$, while remaining essentially unchanged at $D=1\pc$. The rotation is strongest in the paramagnetic (PM) case because the lower magnetic susceptibility places a larger fraction of grains in the $\kRAT$ regime. The angle remains at its pre-SN value during the fast-alignment flare and rotates only as the system enters the disruption-driven polarization dip.

As the SN luminosity declines, radiative precession is slower and eventually becomes subdominant to Larmor precession. The alignment axis then returns to $\Bv$, and the polarization angle relaxes toward its original orientation, completing a polarization-angle reverberation cycle. For clouds at $D\sim0.1$--$1\pc$, the full cycle lasts approximately $200$--$600$ days, with its duration determined by the magnetic properties of the grains: SPM grains complete the cycle within $\sim200$--$300$ days, whereas PM grains at $D=0.1\pc$ remain rotated at the last epoch computed ($\sim600$ days).

A control calculation with RAT-D disabled exhibits no polarization-angle rotation (\paperii). In this case, the high-$J$ grains are not destroyed, and their $\BRAT$ alignment continues to dominate the net polarization. The observable angle jump therefore requires both access to the $\kRAT$ regime and disruption of the dominant $\BRAT$-aligned population; neither transient heating nor enhanced alignment alone can produce it. This interpretation is further supported by the validation case $\bk\parallel\bB$, for which no angle rotation occurs because the two alignment axes coincide in projection (\paperii).

\subsection{Polarization-angle reverberation as a probe of grain magnetism}
\label{sec:reverberation}

The timing of the polarization-angle evolution provides a sensitive diagnostic of grain magnetic properties. For paramagnetic (PM) grains, in which iron atoms are dispersed throughout the silicate matrix, the zero-frequency magnetic susceptibility follows the Curie law, $\chi_{0}\propto T_d^{-1}$. Transient heating therefore reduces the susceptibility, slows Larmor precession, and promotes a switch of the alignment axis from $\Bv$ to $\kv$ on timescales of days to weeks for clouds close to the SN. By contrast, superparamagnetic (SPM) grains containing iron clusters with $\Ncl\gtrsim10^{3}$ atoms maintain an approximately constant effective susceptibility \citep{Hoang.2026}. Their rapid Larmor precession allows $\BRAT$ alignment to persist under substantially stronger transient illumination \citep{Hoang.2026}.

This magnetic dichotomy is directly visible in Figures~\ref{fig:pext_mag_Bk45} and \ref{fig:pem_mag_Bk45}. At the same epoch and cloud distance, PM and SPM grains can occupy different alignment regimes, producing polarization angles that differ by as much as the projected angle $\psi$ between $\Bv$ and $\kv$. The distinction reappears during the recovery phase. As the transient fades, SPM grains return from $\kRAT$ to $\BRAT$ more rapidly than PM grains because of their faster Larmor precession, again producing an angular separation of up to $\psi$ between the two cases.

The complete $\Bv\to\kv\to\Bv$ excursion constitutes a polarization-angle ``reverberation.'' Both its onset and recovery encode grain magnetism, but the recovery phase provides the cleaner diagnostic. The onset depends jointly on the radiation strength and magnetic susceptibility, whereas the restoration of the $\BRAT$ orientation as the transient fades is governed primarily by the Larmor precession rate and hence by the grain iron content. SPM grains should therefore recover their pre-transient polarization angle substantially earlier than PM grains. Multi-epoch polarimetry of a single transient can thus constrain the magnetic response of interstellar grains---information that is largely inaccessible to static polarization measurements.
%Circumstellar dust around evolved massive stars is known to carry metallic iron inclusions \citep{Kemper.2002}, and the iron content changes the susceptibility by up to three orders of magnitude \citep{Hoang.2026}; 

\subsection{Distance dependence of the polarization diagnostics}
\label{sec:diag}

Table~\ref{tab:diag} summarizes the predicted onset times and amplitudes of the polarization signatures in both $V$-band extinction, $\pext(V)$, and $850\mum$ thermal emission, $\pem(850\mum)$, as a function of SN-cloud distance. For nearby clouds with $D\lesssim1\pc$, RAT-D removes the large grains whose $\BRAT$ alignment dominates the polarized signal, allowing the transition from $\Bv$ to $\kv$ to produce an observable polarization-angle rotation. %Rotational disruption also drives a non-monotonic evolution of the total-to-selective extinction ratio, $\RVsn(t_r)$: after an initial rise, $\RVsn$ can decrease by up to $\Delta\RVsn\sim-1$, reaching $\RVsn\simeq3$ at $D\lesssim0.01\pc$ (\paperii).

In general, decreasing $D$ produces earlier and stronger polarization variability because the cloud experiences a more intense radiation field. At $D\gtrsim5\pc$, the response is weaker and slower, with little rotational disruption or polarization-angle rotation, making these distant clouds useful reference environments. Joint optical and submillimeter polarimetry can therefore use the timing and amplitude of the transient response to constrain the location of the illuminated dust, while the polarization-angle evolution probes its magnetic properties.

\begin{table*}[ht]
	\centering
	\caption{Predicted observational diagnostics for dense clouds at different distances from a Type~IIP SN ($\nH=10^{4}\cm^{-3}$, $\Smax=10^{9}\erg\cm^{-3}$, $B=100\,\mu$G).}
	\label{tab:diag}
	\footnotesize
	\setlength{\tabcolsep}{2.5pt}
	\begin{tabular}{lcccccc}
		\hline\hline
		$D$ (pc) & Sublimation & RAT-D onset & $\kRAT$ (PM) & $\pext(V)/N_{\rm H}(\%/10^{21}\cm^{-2})$ peak & $\pem$ peak &
		$\lambda_{\max}$ shift ($\mum$) \\
		\hline
		0.01 & $\sim 2$\,d & $<1$\,d       & yes (until subl.) & $\sim 1.45\to 0$  & $\sim 14.2\%\to 0$  & $0.8\to\sim 0.11$ (until subl.) \\
		0.1  & none        & $\sim 1.5$\,d   & $\sim 1.3-\gtrsim600$ d & $\sim 1.34$   & $\sim 13.4\%$   & $0.8\to\sim 0.13$ \\
		1    & none        & $\sim 20$\,d  & $\sim 20-400$ d & $\sim 1.35$   & $\sim 13.3\%$   & $0.8\to\sim 0.14$ \\
		5    & none        & marginal      & no & $\sim 1.34$ & $\sim 12.9\%$ & $0.8\to\sim 0.15$ \\
		10   & none        & none          & no & $\sim 1.32$& $\sim 12.2\%$& $0.8\to\sim 0.16$ \\
		20   & none        & none          & no & $\sim 1.29$& $\sim 11.3\%$& $0.8\to\sim 0.19$ \\
		\hline
	\end{tabular}
	\tablecomments{$\pext(V)$: $V$-band extinction polarization of background starlight through the cloud; $\pem$: $850\mum$ emission polarization (quiescent value $\simeq 7.8\%$; the quiescent $\pext(V)/N_{\rm H}\simeq0.83$). $\lambda_{\max}$ shift: blueward drift of the extinction-polarization peak wavelength from its quiescent value to the post-flare plateau (Fig.~\ref{fig:lambda_peak_mag_Bk45}). The $\kRAT$ (PM) column indicates whether the $\BRAT\to\kRAT$ switch occurs for PM grains; SPM grains retain $\BRAT$ alignment longer and also relax to $\BRAT$ faster (Section~\ref{sec:reverberation}).}
\end{table*}

\section{Discussion and conclusions}
\label{sec:discussion}
\subsection{Fast alignment and rotational disruption: time-domain dust physics}
\label{sec:fundamental}

We identify a distinctive signature of dynamical dust physics in transient environments: the coupled effects of \emph{fast alignment and rotational disruption}, manifested as a polarization flare followed by a polarization dip in both extinction and thermal emission (Section~\ref{sec:flare}). The flare probes the grain-alignment timescale and the fraction of grains driven to high-$J$ attractors, whereas the dip measures their susceptibility to centrifugal disruption. The amplitudes and timescales of these two phases depend differently on the radiation intensity, parameterized by the transient-cloud distance $D$, and on the grain tensile strength $\Smax$, allowing the underlying processes to be disentangled. The accompanying blueward drift of the extinction-polarization peak wavelength, $\lambda_{\max}$ (Section~\ref{sec:lambdamax}), provides an independent record of the evolving alignment and disruption thresholds. Unlike polarization-amplitude diagnostics, $\lambda_{\max}$ is not degenerate with the inclination of the magnetic field. During fast alignment, the polarization angle retains its pre-transient orientation. It rotates only after rotational disruption removes the large grains aligned at high-$J$ attractors through $\BRAT$, causing the dominant alignment axis to switch from $\Bv$ to $\kv$. Fast alignment and rotational disruption are fundamental predictions of the RAT paradigm \citep{Hoangetal.2019,Hoang.2025}, but quiescent environments constrain only their final outcomes. By switching an intense radiation field on and off over days to months, a transient reveals these processes \emph{in action} and helps break the degeneracies among the grain-size distribution, alignment efficiency, and magnetic properties that limit static modeling.

To date, no time-resolved monitoring of thermal dust polarization from transient-illuminated dust has been conducted. However, the predicted signatures are accessible to existing facilities (Section~\ref{sec:prospects}). Their detection would directly test the dynamical core of the RAT paradigm, whereas robust nondetections would place stringent constraints on---or potentially challenge---its predictions.

\subsection{Dust physical memory effects and fossil imprints in observables: a hierarchy of transient-driven dust memory}
\label{sec:memory}
Our results based on the \TransRAT\ framework reveal a strong asymmetry between how rapidly dust \emph{responds} to an SN transient and how slowly it \emph{forgets}. The response is governed by the duration of the radiation pulse---days to months---and the dust location, whereas the return toward the pre-SN state occurs over a characteristic relaxation timescale, $\tau_{\rm relax}$. Here we discuss key {\it physical memory effects} caused by transient-irradiation on dust and {\it fossil imprints} in observables such as dust extinction and polarization.

\emph{Alignment-axis memory}: As the transient radiation fades, anomalous $\kRAT$ polarization angles rapidly revert to the $\BRAT$ orientation because of fast Larmor precession, with $\tau_{\rm relax}\sim\tau_{\rm Lar}$. Polarization-angle anomalies are therefore primarily real-time diagnostics, observable only during and shortly after the radiation pulse.

\emph{Fast-alignment memory}: Fast alignment produces an enhanced population of grains at high-$J$ attractors and, consequently, an elevated polarization degree. This enhancement decays only as gas collisions damp and randomize the grain angular momentum. Following the transient, the angular velocity decreases from its maximum value according to
\begin{equation}
	\Omega(t)=\Omega_{\max}\exp\left(-\frac{t}{t_{\rm gas}}\right).
\end{equation}
The corresponding relaxation time can be estimated as the time required for the angular velocity to decrease from $\Omega_{\max}$ to its thermal value $\Omega_T$:
\begin{equation}
	\tau_{\rm relax}\sim
	\ln\left(\frac{\Omega_{\max}}{\Omega_T}\right)t_{\rm gas}
	\sim 1-10\,t_{\rm gas},\label{eq:trelax_gas}
\end{equation}
for $\Omega_{\max}/\Omega_{T}\sim 3-10^{4}$.

This timescale defines the lifetime of the fast-alignment fossil. Smaller grains lose their transient-induced fast alignment first because they attain lower $\Omega_{\max}$ and have shorter gas-damping times. The grain population eventually returns to its pre-transient alignment state, in which only grains larger than $a_{\rm ali}^{\rm ISRF}$ remain aligned by the interstellar radiation field.

\emph{Disruption memory}: In contrast to alignment, RAT-D is irreversible. The resulting modification of the grain-size distribution and reduction in the total-to-selective extinction, $\RVsn$, persist until grain growth replenishes the disrupted population (\paperii). The relevant relaxation time is therefore the regrowth timescale, $\tau_{\rm relax}\sim\tau_{\rm regrowth}\gtrsim{100\rm kyr}$ even in dense gas, making disruption memory the most durable fossil signature. Around CCSNe, pre-existing dust should consequently retain a greater abundance of small grains than it had before the explosion.

\emph{Shape-deformation memory}: Finally, prior to disruption, rapid rotation can deform grains into increasingly oblate shapes, with larger axial ratios \citep{Reissl.2024}. The resulting deformed grain shape, together with the associated enhancement in intrinsic polarization efficiency, can persist until subsequent gas accretion or grain–grain collisions reshape or replace the grain.

We have extended our calculations to $10\yr$, far beyond the duration of the SN radiation pulse, and found that the enhanced-alignment and disruption signatures remain similar to those at the final epochs shown in Figures~\ref{fig:pext_mag_Bk45}--\ref{fig:pem_mag_Bk45_R1-10}. When combined with light-travel-time effects, these persistent signatures form an expanding shell of modified dust: a ``polarization echo'' and a ``disruption echo,'' analogous to a scattered-light echo but persisting locally for $\tau_{\rm relax}$ rather than only for the duration of the flash. The \TransRAT\ framework can therefore be tested both in real time and \emph{archaeologically} through the fossil record of past transients (Section~\ref{sec:prospects}).

\subsection{Probing pristine magnetic fields before shock arrival}

Pre-shock magnetic fields provide essential information about transient progenitor environments and the initial conditions for cosmic-ray acceleration. However, radio synchrotron polarization from SNRs probes the field only after it has been compressed and amplified by the shock. A radiation front propagating at $c$ reaches a clump at distance $D>0.1\pc$ after
\begin{equation}
	t_{\rm rad}\simeq0.33\left(\frac{D}{0.1\pc}\right)\yr,
\end{equation}
whereas a blast wave traveling at $v_{\rm sh}\sim10^{4}\km\,{\rm s}^{-1}$ arrives only after
\begin{equation}
	t_{\rm sh}\simeq9.8\left(\frac{D}{0.1\pc}\right)
	\left(\frac{v_{\rm sh}}{10^{4}\km\,{\rm s}^{-1}}\right)^{-1}\yr.
\end{equation}
The resulting interval provides a unique window in which the polarization flare is governed by $\BRAT$ alignment with the undisturbed magnetic field. During this period, the polarization angle of superparamagnetic grains---and of paramagnetic grains at sufficiently early times---maps the plane-of-sky orientation of the pristine field in the illuminated clump. Once the blast wave reaches the same region, comparison with its post-shock radio synchrotron polarization can isolate the magnetic-field compression, reorientation, and amplification produced by the shock.

\subsection{Observational prospects}
\label{sec:prospects}
We now discuss two complementary strategies for testing our predictions: real-time polarimetric monitoring of transient-illuminated dust and archaeological searches for fossil imprints around past transients.
\subsubsection{Real-time polarimetric tests}
\label{sec:prospects_realtime}

Galactic and Local Group transients offer the most direct tests because their illuminated dust structures may be spatially resolved. Nearby CCSNe discovered by the Vera C. Rubin Observatory's Legacy Survey of Space and Time (LSST), whose ten-year survey began in mid-2026 \citep{Ivezic.2019}, can be monitored for delayed polarized emission from off-axis clouds over the following years. For a characteristic cloud distance $D\sim1\pc$, the delay is $t_{\rm delay} \simeq3.3(D/{\rm pc})\yr$, with the precise delay depending on the cloud geometry. The associated infrared echo---in which dust emission is enhanced by a factor of $\sim U$ and the spectral energy distribution shifts toward the mid- and far-infrared---provides a natural trigger for polarimetric follow-up. Its total intensity is within reach of JWST, while the far-infrared probe PRIMA could provide polarimetric imaging in the early 2030s.

At submillimeter wavelengths, SCUBA-2/POL-2 on the JCMT can search for the predicted $850\mum$ polarization flare, while ALMA can resolve circumstellar shells and dense protostellar cores. The proposed 50-m AtLAST would extend such monitoring to wider fields. Proposed UV spectropolarimetry missions, including PUFFINS \citep{Anche.2025}, would be particularly well suited to detecting polarization-angle reverberation and the flare--dip sequence.

For the much larger population of extragalactic SNe, individual illuminated clouds will generally remain unresolved, and the observed signal will be integrated along the line of sight. The predicted response can nevertheless be tested through temporal changes in the optical/NIR polarization degree, position angle, and peak wavelength $\lambda_{\max}$ of the SN light using existing spectropolarimeters such as VLT/FORS2 and Keck/LRISp. In this occulting geometry, $\kv$ is directed along the line of sight, so grains aligned through $\kRAT$ produce no dichroic polarization. For nearby clouds with $D\lesssim1\pc$, the onset of the $\BRAT\to\kRAT$ transition therefore causes $\pext$ of the SN light to decrease at an approximately constant position angle. The magnitude of this decrease is set by the fraction of grains undergoing $\kRAT$ alignment: it is largest for paramagnetic grains and becomes progressively smaller with increasing magnetic susceptibility. Thus, instead of the $45^{\circ}$ rotation predicted for the off-axis geometry in Figure~\ref{fig:pext_mag_Bk45}, the occulting geometry produces a characteristic drop in polarization amplitude. Time-variable polarization has already been reported for SN~2014J in M82 \citep{Patat:2015bb,Hoang.2017} and for several CCSNe with imaging-polarimetric monitoring. Although such variability has commonly been attributed to dust in nearby clouds, the \TransRAT\ framework provides a quantitative diagnostic for identifying transient-driven alignment and disruption.

Because of the memory hierarchy described in Section~\ref{sec:memory}, the enhanced-alignment and disruption signatures persist beyond the locally bright phase. After the transient fades, the polarization fraction remains elevated while the position angle relaxes toward its pre-SN value. By contrast, the polarized emission intensity, governed by the evolving dust temperature $T_d(t)$, declines with the illuminating source. These distinct temporal behaviors provide a means of separating transient dust heating from the longer-lived effects of grain alignment and disruption.

\subsubsection{Archaeological tests of physical memory effects in molecular clouds near young SNRs}
\label{sec:prospects_fossil}

The hierarchy of relaxation times of physical memory effects enables an archaeological test of the \TransRAT\ framework. For a cloud at distance $D$ from a supernova, the relaxation clock begins only when the radiation pulse reaches the cloud. Its post-illumination age is therefore
\begin{equation}
	t_{\rm post}=t_{\rm age}-\frac{D}{c},
\end{equation}
and a given memory channel survives when
\begin{equation}
	0<t_{\rm post}<\tau_{\rm relax},
\end{equation}
where the lower bound ensures that the cloud has already been illuminated. For the fast-alignment memory, with $\tau_{\rm relax}\sim10\,t_{\rm gas}$, clouds satisfying this condition should retain an enhanced polarization degree and a blueshifted $\lambda_{\max}$, together with the reduced $\RVsn$ produced by RAT-D. Around older remnants, the alignment enhancement should have relaxed toward the ambient polarization baseline, while the reduced $\RVsn$ and part of the $\lambda_{\max}$ blueshift persist as irreversible disruption memory.

The fast-alignment relaxation time is approximately (see Eq. \ref{eq:trelax_gas})
\begin{equation}
	\tau_{\rm relax}\sim10\,t_{\rm gas}
	\sim10^{3}\text{--}10^{4}
	\left(\frac{\nH}{10^{4}\cm^{-3}}\right)^{-1}\yr.
\end{equation}

Clumps with $\nH\lesssim10^{3}\cm^{-3}$ around the youngest Galactic remnants---including Cas~A ($\sim350\yr$), Tycho's SNR ($\sim450\yr$), and SN~1006 ($\sim10^{3}\yr$)---could therefore retain fast-alignment memory today. The still longer relaxation time in diffuse gas implies that a substantial fraction of the illuminated volume outside the blast wave may preserve enhanced-alignment and RAT-D signatures. At a fixed remnant age, clouds at larger $D$ were illuminated more recently and should consequently be less relaxed, provided that the radiation pulse was sufficiently intense to modify their grains. This timing effect competes with the decrease in radiation strength with distance.

The observational test is fundamentally differential. Clouds around SNRs should be selected within the illuminated volume but outside the blast radius, thereby minimizing contamination from direct shock processing, and compared with matched, unexposed control clouds. The samples should be stratified by remnant age, cloud density, and distance from the explosion. Because the polarization degree also depends on magnetic-field inclination, the geometry-independent quantities $\lambda_{\max}$ and $\RVsn$ provide the cleanest diagnostics for individual clouds; an enhancement in polarization degree is more robustly established statistically across a cloud sample. SNR--molecular-cloud associations are already cataloged \citep{Zhou.2023}, and the predicted fossil signatures are accessible through wide-field starlight-polarization surveys such as PASIPHAE \citep{Tassis.2018}, combined with Gaia-based three-dimensional dust maps. The newly commissioned wide-field imaging polarimeter DragonflyPol \citep{Tahani.2026} provides a complementary probe whose sensitivity to low-surface-brightness polarized emission is well matched to the degree-scale environments of historical SNRs. This archaeological approach is therefore the most practical near-term test of the framework: it requires neither a serendipitous nearby transient nor a rapid polarimetric trigger, and statistical samples can be assembled using facilities already in operation.

Because the \TransRAT\ framework accepts an arbitrary light curve, the same real-time and archaeological diagnostics extend naturally to other transient sources, with onset times, amplitudes, and relaxation histories rescaled according to the radiation field and environment. Relevant applications include SNe~Ia viewed through foreground clouds, GRB afterglows embedded in host molecular clouds, novae and fast blue optical transients surrounded by dense circumstellar material, and TDEs within dusty tori. The latter are particularly promising because recent optical polarimetry has revealed high and rapidly variable polarization \citep{Floris.2025,Koljonen.2025,Uno.2025}. The framework also applies to flares from active galactic nuclei and episodic outbursts from young stellar objects (\paperii).
\subsection{Summary}

We apply the time-domain \TransRAT\ framework (\paperii) to a molecular cloud illuminated by a Type~IIP SN and predict four polarization signatures whose amplitudes and timescales vary systematically with the SN-cloud distance (Table~\ref{tab:diag}). First, at $D\lesssim1\pc$, both thermal dust polarization and extinction polarization rise rapidly in a \emph{polarization flare} and subsequently decline in a \emph{polarization dip} as RAT-D destroys the large aligned grains. Second, rapid alignment of progressively smaller grains drives a blueward drift of the extinction-polarization peak wavelength, $\lambda_{\max}$. This drift directly traces the minimum fast-alignment size without geometric degeneracies and persists after the SN fades (Sec.~\ref{sec:lambdamax}). Third, the $\BRAT\to\kRAT$ transition produces an abrupt polarization-angle rotation---reaching $\simeq45^{\circ}$ for our fiducial geometry with $\psi=45^{\circ}$ at $D\lesssim0.1\pc$, and up to $\sim90^{\circ}$ in general. This rotation depends on disruption physics and disappears in the no-RAT-D control. Fourth, the subsequent polarization-angle reverberation provides a direct measure of grain iron content through its timing, distinguishing paramagnetic from superparamagnetic dust.

Beyond these real-time signatures, the transient leaves enduring \emph{fossil imprints}. The enhanced polarization degree and blueshifted $\lambda_{\max}$ persist for a relaxation time $\tau_{\rm relax}\sim10\,t_{\rm gas}$ after the radiation fades, whereas the reduced $\RVsn$ and part of the $\lambda_{\max}$ blueshift survive even longer as an irreversible memory of grain disruption (Sec.~\ref{sec:memory}). Together, these signatures reveal fast-alignment and rotational-disruption physics that only a transient radiation field can expose in real time, while probing pristine pre-shock magnetic fields and grain magnetism years before the shock arrives. The polarization-angle rotation provides the cleanest real-time discriminator because it is difficult for alternative mechanisms to reproduce, whereas fossil signatures in clouds near young SNRs offer the most practical near-term test because they require no real-time trigger (Sec.~\ref{sec:prospects_fossil}). Time-resolved polarimetric monitoring of SN-illuminated dust, complemented by spatially resolved polarimetric mapping of clouds around Galactic and nearby SNRs, can directly test this physics and establish dust polarization as a new messenger of time-domain astrophysics.

\begin{acknowledgments}
This work was supported by the major research project (No. 2026183200) from Korea Astronomy and Space Science Institute (KASI) funded by the Ministry of Science and ICT (MSIT). I am grateful to Alex Lazarian for insightful discussions on grain alignment physics. This work was partially funded by a grant from the Simons Foundation (SFI-MPS-T-Institutes-00021894, HTN). We thank the ICISE staff for excellent support and hospitality. Claude \citep{claude} was used to assist with code development and ChatGPT \citep{gpt} with polishing the manuscript.

\end{acknowledgments}

\software{Python, \texttt{DustPOL\_py} \citep{Tram.2023}, ChatGPT \citep{gpt}, Claude \citep{claude}.}

%\bibliographystyle{aasjournalv7}
%\bibliography{/Users/thiemhoang/Dropbox/Papers3/cites_paperApJ,ms_extra}
%\bibliography{cites_paperApJ,ms_extra}% <-- switch to this (local) form for submission
\bibliography{ms.bbl}

\begin{thebibliography}{}
\expandafter\ifx\csname natexlab\endcsname\relax\def\natexlab#1{#1}\fi
\providecommand{\url}[1]{\href{#1}{#1}}
\providecommand{\dodoi}[1]{doi:~\href{http://doi.org/#1}{\nolinkurl{#1}}}
\providecommand{\doeprint}[1]{\href{http://ascl.net/#1}{\nolinkurl{http://ascl.net/#1}}}
\providecommand{\doarXiv}[1]{\href{https://arxiv.org/abs/#1}{\nolinkurl{https://arxiv.org/abs/#1}}}

% type= article
\bibitem[{R.~M. {Anche} {et~al.}(2025){Anche}, {Kang}, {Van Gorkom}, {Vargas},
  {Chung}, {Kupinski}, {Andersson}, {Clayton}, {Douglas}, {Hamden}, {Hoang}, \&
  et~al.}]{Anche.2025}
{Anche}, R.~M., {Kang}, H., {Van Gorkom}, K., {et~al.} 2025,
  \bibinfo{title}{{Optical design and polarimetric performance of a SmallSat UV
  polarimeter to study interstellar dust: PUFFINS},} Proc. SPIE, 13615,
  arXiv:2509.02994

% type= article
\bibitem[{B.-G. Andersson {et~al.}(2015)Andersson, Lazarian, \&
  Vaillancourt}]{Andersson.2015}
Andersson, B.-G., Lazarian, A., \& Vaillancourt, J.~E. 2015,
  \bibinfo{title}{{Interstellar Dust Grain Alignment},} \araa, 53, 501

% type= misc
\bibitem[{ {Anthropic}(2026){Anthropic}}]{claude}
{Anthropic}. 2026, {Claude Code},, \url{https://claude.com/claude-code}

% type= article
\bibitem[{R. {Dastidar} {et~al.}(2018){Dastidar}, {Misra}, {Hosseinzadeh},
  {Pastorello}, {Pumo}, {Valenti}, \& et~al.}]{Dastidar.2018}
{Dastidar}, R., {Misra}, K., {Hosseinzadeh}, G., {et~al.} 2018,
  \bibinfo{title}{{SN 2015ba: a Type IIP supernova with a long plateau},}
  \mnras, 479, 2421, \dodoi{10.1093/mnras/sty1634}

% type= article
\bibitem[{B. Dinçel {et~al.}(2026)Dinçel, Paylı, Yerli, Ankay, Neuhäuser,
  Mugrauer, Sheth, Buder, Hüttel, Edelmann, Michel, \& Bätz}]{Dincel.2026}
Dinçel, B., Paylı, G., Yerli, S.~K., {et~al.} 2026, \bibinfo{title}{{Massive
  runaway star HD 254577: The pre-supernova binary companion to the progenitor
  of the supernova remnant IC 443},} Astronomy \& Astrophysics, 707, A50,
  \dodoi{10.1051/0004-6361/202556086}

% type= article
\bibitem[{A.~Z. Dolginov \& I.~G. Mitrofanov(1976)Dolginov \&
  Mitrofanov}]{Dolginov.1976}
Dolginov, A.~Z., \& Mitrofanov, I.~G. 1976, \bibinfo{title}{{Orientation of
  cosmic dust grains},} Ap\&SS, 43, 291

% type= article
\bibitem[{B.~T. Draine \& J.~C. Weingartner(1997)Draine \&
  Weingartner}]{DraineWein.1997}
Draine, B.~T., \& Weingartner, J.~C. 1997, \bibinfo{title}{{Radiative Torques
  on Interstellar Grains. II. Grain Alignment},} \apj, 480, 633

% type= article
\bibitem[{A. Floris {et~al.}(2025)Floris, Liodakis, Koljonen, Lindfors,
  Agís-Gonzalez, Paggi, Blinov, Nilsson, Agudo, Charalampopoulos, Teodori,
  Pedrosa, Otero-Santos, Piirola, Newsome, \& Velzen}]{Floris.2025}
Floris, A., Liodakis, I., Koljonen, K. I.~I., {et~al.} 2025,
  \bibinfo{title}{{Polarimetric diversity in tidal disruption events:
  Comparative study of low-polarised sources with AT2020mot},} Astronomy \&
  Astrophysics, 703, A81, \dodoi{10.1051/0004-6361/202555626}

% type= article
\bibitem[{N.~C. Giang {et~al.}(2020)Giang, Hoang, \& Tram}]{Gianghoang.2020}
Giang, N.~C., Hoang, T., \& Tram, L.~N. 2020, \bibinfo{title}{{Time-varying
  Extinction, Polarization, and Colors of Type Ia Supernovae due to Rotational
  Disruption of Dust Grains},} \apj, 888, 93

% type= article
\bibitem[{B.~S. Hensley \& B.~T. Draine(2023)Hensley \&
  Draine}]{DraineHensley.2023}
Hensley, B.~S., \& Draine, B.~T. 2023, \bibinfo{title}{{The Astrodust+PAH
  Model: A Unified Description of the Extinction, Emission, and Polarization
  from Dust in the Diffuse Interstellar Medium},} \apj, 948, 55

% type= article
\bibitem[{J.~T. {Hinkle} {et~al.}(2025){Hinkle}, {Shappee}, {Auchettl}, \&
  et~al.}]{Hinkle.2025}
{Hinkle}, J.~T., {Shappee}, B.~J., {Auchettl}, K., \& et~al. 2025,
  \bibinfo{title}{{The most energetic transients: Tidal disruptions of
  high-mass stars},} Science Advances, 11, eadt0074,
  \dodoi{10.1126/sciadv.adt0074}

% type= article
\bibitem[{T. Hoang(2017)Hoang}]{Hoang.2017}
Hoang, T. 2017, \bibinfo{title}{{Properties and Alignment of Interstellar Dust
  Grains toward Type Ia Supernovae with Anomalous Polarization Curves},} \apj,
  836, 13

% type= article
\bibitem[{T. Hoang(2025)Hoang}]{Hoang.2025}
Hoang, T. 2025, \bibinfo{title}{{Toward a General Theory of Grain Alignment and
  Disruption by Radiative Torques and Magnetic Relaxation},} The Astrophysical
  Journal, 994, 115, \dodoi{10.3847/1538-4357/ae0a1a}

% type= article
\bibitem[{T. Hoang(2026{\natexlab{a}})Hoang}]{Hoang.2026}
Hoang, T. 2026{\natexlab{a}}, \bibinfo{title}{{Effective Magnetic
  Susceptibility of Dust Grains with Superparamagnetic Inclusions and
  Implications},} arXiv:2602.14101

% type= article
\bibitem[{T. Hoang(2026{\natexlab{b}})Hoang}]{Hoang.2026klf}
Hoang, T. 2026{\natexlab{b}}, \bibinfo{title}{{Time-Domain Dust Astrophysics.
  II. TransRAT: Time-Dependent Grain Alignment and Disruption by Cosmic
  Transients and Their Observational Signatures},} arXiv:2609.09555

% type= article
\bibitem[{T. Hoang {et~al.}(2020)Hoang, Giang, \& Tram}]{Hoanggiang.2020}
Hoang, T., Giang, N.~C., \& Tram, L.~N. 2020, \bibinfo{title}{{Gamma-Ray Burst
  Afterglows: Time-varying Extinction, Polarization, and Colors due to
  Rotational Disruption of Dust Grains},} \apj, 895, 16

% type= article
\bibitem[{T. Hoang \& A. Lazarian(2008)Hoang \& Lazarian}]{HoangLaz.2008}
Hoang, T., \& Lazarian, A. 2008, \bibinfo{title}{{Radiative torque alignment:
  essential physical processes},} \mnras, 388, 117

% type= article
\bibitem[{T. Hoang \& A. Lazarian(2014)Hoang \& Lazarian}]{Hoangetal.2014}
Hoang, T., \& Lazarian, A. 2014, \bibinfo{title}{{Grain alignment by radiative
  torques in special conditions and implications},} \mnras, 438, 680

% type= article
\bibitem[{T. Hoang \& A. Lazarian(2016)Hoang \& Lazarian}]{HoangLaz.2016}
Hoang, T., \& Lazarian, A. 2016, \bibinfo{title}{{A Unified Model of Grain
  Alignment: Radiative Alignment of Interstellar Grains with Magnetic
  Inclusions},} \apj, 831, 159

% type= article
\bibitem[{T. Hoang {et~al.}(2019)Hoang, Tram, Lee, \& Ahn}]{Hoangetal.2019}
Hoang, T., Tram, L.~N., Lee, H., \& Ahn, S.-H. 2019,
  \bibinfo{title}{{Rotational disruption of dust grains by radiative torques in
  strong radiation fields},} natas, 3, 766

% type= article
\bibitem[{T. Hoang {et~al.}(2021)Hoang, Tram, Lee, Diep, \& Ngoc}]{Hoang.2021}
Hoang, T., Tram, L.~N., Lee, H., Diep, P.~N., \& Ngoc, N.~B. 2021,
  \bibinfo{title}{{Grain Alignment and Disruption by Radiative Torques in Dense
  Molecular Clouds and Implication for Polarization Holes},} \apj, 908, 218,
  \dodoi{10.3847/1538-4357/abd54f}

% type= article
\bibitem[{T. {Hoang} {et~al.}(2022){Hoang}, {Tram}, {Minh Phan}, {Giang},
  {Phuong}, \& {Dieu}}]{Hoangetal.2022}
{Hoang}, T., {Tram}, L.~N., {Minh Phan}, V.~H., {et~al.} 2022,
  \bibinfo{title}{{On Internal and External Alignment of Dust Grains in
  Protostellar Environments},} \aj, 164, 248, \dodoi{10.3847/1538-3881/ac9af5}

% type= article
\bibitem[{{\v Z}. {Ivezi{\'c}} {et~al.}(2019){Ivezi{\'c}}, {Kahn}, {Tyson}, \&
  et~al.}]{Ivezic.2019}
{Ivezi{\'c}}, {\v Z}., {Kahn}, S.~M., {Tyson}, J.~A., \& et~al. 2019,
  \bibinfo{title}{{LSST: From Science Drivers to Reference Design and
  Anticipated Data Products},} \apj, 873, 111

% type= article
\bibitem[{K.~I.~I. Koljonen {et~al.}(2025)Koljonen, Nilsson, Liodakis, \&
  Lindfors}]{Koljonen.2025}
Koljonen, K. I.~I., Nilsson, K., Liodakis, I., \& Lindfors, E. 2025,
  \bibinfo{title}{{Optical polarization properties of the closest tidal
  disruption event AT 2023clx indicate origin from tidal stream shocks},}
  Monthly Notices of the Royal Astronomical Society, 542, 2238,
  \dodoi{10.1093/mnras/staf1340}

% type= article
\bibitem[{A. Lazarian \& T. Hoang(2007)Lazarian \& Hoang}]{LazHoang.2007}
Lazarian, A., \& Hoang, T. 2007, \bibinfo{title}{{Radiative torques: analytical
  model and basic properties},} \mnras, 378, 910

% type= article
\bibitem[{A. Lazarian \& T. Hoang(2019)Lazarian \& Hoang}]{LazHoang.2019}
Lazarian, A., \& Hoang, T. 2019, \bibinfo{title}{{Magnetic Properties of Dust
  Grains, Effect of Precession, and Radiative Torque Alignment},} \apj, 883,
  122

% type= article
\bibitem[{H. Lee {et~al.}(2020)Lee, Hoang, Le, \& Cho}]{LeeHoang.2020}
Lee, H., Hoang, T., Le, N., \& Cho, J. 2020, \bibinfo{title}{{Physical Model of
  Dust Polarization by Radiative Torque Alignment and Disruption and
  Implications for Grain Internal Structures},} \apj, 896, 44

% type= misc
\bibitem[{ {OpenAI}(2025){OpenAI}}]{gpt}
{OpenAI}. 2025, {ChatGPT},, \url{https://chatgpt.com}

% type= article
\bibitem[{F. Patat {et~al.}(2015)Patat, Taubenberger, Cox, Baade, Clocchiatti,
  H{\"o}flich, Maund, Reilly, Spyromilio, Wang, Wheeler, \&
  Zelaya}]{Patat:2015bb}
Patat, F., Taubenberger, S., Cox, N. L.~J., {et~al.} 2015,
  \bibinfo{title}{{Properties of extragalactic dust inferred from linear
  polarimetry of Type Ia Supernovae},} \aa, 577, A53

% type= article
\bibitem[{K. Pattle \& L. Fissel(2019)Pattle \& Fissel}]{PattleFissel.2019}
Pattle, K., \& Fissel, L. 2019, \bibinfo{title}{{Submillimeter and Far-infrared
  Polarimetric Observations of Magnetic Fields in Star-Forming Regions},}
  Frontiers in Astronomy and Space Sciences, 6, 15

% type= article
\bibitem[{W.~T. Reach {et~al.}(2024)Reach, Tram, DeWitt, Lesaffre, Godard, \&
  Gusdorf}]{Reach.2024}
Reach, W.~T., Tram, L.~N., DeWitt, C., {et~al.} 2024,
  \bibinfo{title}{{Supernova Shocks in Molecular Clouds: Shocks Driven into
  Dense Cores in IC 443 and 3C 391},} The Astrophysical Journal, 977, 149,
  \dodoi{10.3847/1538-4357/ad8d59}

% type= article
\bibitem[{S. Reissl {et~al.}(2024)Reissl, Nguyen, Jordan, \&
  Klessen}]{Reissl.2024}
Reissl, S., Nguyen, P., Jordan, M.~L., \& Klessen, S.~R. 2024,
  \bibinfo{title}{{The rotational disruption of porous dust aggregates from ab
  initio kinematic calculations},} Astronomy \& Astrophysics,
  \dodoi{10.1051/0004-6361/202346068}

% type= article
\bibitem[{P. Slane {et~al.}(2015)Slane, Bykov, Ellison, Dubner, \&
  Castro}]{slane.2015supernova-c88}
Slane, P., Bykov, A., Ellison, D.~C., Dubner, G., \& Castro, D. 2015,
  \bibinfo{title}{Supernova Remnants Interacting with Molecular Clouds: X-Ray
  and Gamma-Ray Signatures,} Space Science Reviews, 188, 187,
  \dodoi{10.1007/s11214-014-0062-6}

% type= article
\bibitem[{M. {Tahani} {et~al.}(2026){Tahani}, {Hollberg}, {Akitaya}, {Kim},
  {Lokhorst}, {Abraham}, {van Dokkum}, \& et~al.}]{Tahani.2026}
{Tahani}, M., {Hollberg}, L., {Akitaya}, H., {et~al.} 2026,
  \bibinfo{title}{{DragonflyPol: Wide-Field Optical Linear Polarimetry with the
  Dragonfly Telephoto Array (Instrument Description and Commissioning)},} arXiv
  e-prints, arXiv:2607.14258

% type= article
\bibitem[{K. {Tassis} {et~al.}(2018){Tassis}, {Ramaprakash}, {Readhead}, \&
  et~al.}]{Tassis.2018}
{Tassis}, K., {Ramaprakash}, A.~N., {Readhead}, A.~C.~S., \& et~al. 2018,
  \bibinfo{title}{{PASIPHAE: A high-Galactic-latitude, high-accuracy
  optopolarimetric survey},} arXiv e-prints, arXiv:1810.05652

% type= article
\bibitem[{L.~N. Tram \& T. Hoang(2022)Tram \& Hoang}]{TramHoang.2022}
Tram, L.~N., \& Hoang, T. 2022, \bibinfo{title}{{Recent progress in theory and
  observational study of dust grain alignment and rotational disruption in
  star-forming regions},} Frontiers in Astronomy and Space Sciences, 9, 923927,
  \dodoi{10.3389/fspas.2022.923927}

% type= article
\bibitem[{L.~N. Tram {et~al.}(2023)Tram, Bonne, Hu, Lopez-Rodriguez, Guerra,
  Lesaffre, Gusdorf, Hoang, Lee, Lazarian, Andersson, Coudé, Soam, Vacca, Lee,
  \& Gordon}]{Tram.2023}
Tram, L.~N., Bonne, L., Hu, Y., {et~al.} 2023, \bibinfo{title}{{SOFIA
  Observations of 30 Doradus. II. Magnetic Fields and Large-scale Gas
  Kinematics},} The Astrophysical Journal, 946, 8,
  \dodoi{10.3847/1538-4357/acaab0}

% type= article
\bibitem[{K. Uno {et~al.}(2025)Uno, Maeda, Nagao, Leloudas, Charalampopoulos,
  Mattila, Aoki, Taguchi, Kawabata, Moldon, Pérez-Torres, Pursiainen, \&
  Reynolds}]{Uno.2025}
Uno, K., Maeda, K., Nagao, T., {et~al.} 2025,
  \bibinfo{title}{{Spectropolarimetry of A Nuclear Transient AT2023clx:
  Revealing The Geometrical Alignment between The Transient Outflow and The
  Nuclear Dusty Region},} arXiv

% type= article
\bibitem[{X. Zhou {et~al.}(2023)Zhou, Su, Yang, Chen, Sun, Jiang, Wang, Wang,
  Zhang, Xu, Yan, Yuan, Chen, Ao, \& Ma}]{Zhou.2023}
Zhou, X., Su, Y., Yang, J., {et~al.} 2023, \bibinfo{title}{{A Systematic Study
  of Associations between Supernova Remnants and Molecular Clouds},} The
  Astrophysical Journal Supplement Series, 268, 61,
  \dodoi{10.3847/1538-4365/acee7f}

\end{thebibliography}
\end{document}